\documentclass[aps,prl,preprint,]{revtex4}

\def\ben{\begin{enumerate}} \def\een{\end{enumerate}}
\def\beq{\begin{equation}} \def\eeq{\end{equation}}
\def\bea{\begin{eqnarray}} \def\eea{\end{eqnarray}}
\def\beann{\begin{eqnarray*}} \def\eeann{\end{eqnarray*}}
\def\beasn{\begin{sneqnarray}} \def\eeasn{\end{sneqnarray}}
\begin{document}

\title{Peter Bergmann and the invention of constrained Hamiltonian dynamics}
\author{D. C. Salisbury}
\date{July 23, 2006}

\affiliation{Department of Physics,
Austin College, Sherman, Texas 75090-4440, USA\\E-mail:
dsalisbury@austincollege.edu}

\begin{abstract}
Peter Bergmann was a co-inventor of the algorithm for converting singular Lagrangian models into a constrained Hamiltonian dynamical formalism. This talk focuses on the work of Bergmann and his collaborators in the period from 1949 to 1951.
\end{abstract}


\maketitle

\section{Introduction}
It has always been the practice of those of us associated with the Syracuse ``school'' to identify the algorithm for constructing a canonical phase space description of singular Lagrangian systems as the Dirac-Bergmann procedure. I learned the procedure as a student of Peter Bergmann - and I should point out that he never employed that terminology. Yet it was clear from the published record at the time in the 1970's that his contribution was essential. Constrained Hamiltonian dynamics constitutes the route to canonical quantization of all local gauge theories, including not only conventional general relativity, but also grand unified theories of elementary particle interaction, superstrings and branes. Given its importance and my suspicion that Bergmann has never received adequate recognition from the wider community for his role in the development of the technique, I have long intended to explore this history in depth.  The following is merely a tentative first step in which I will focus principally on the work of Peter Bergmann and his collaborators in the late 1940's and early 1950's, indicating where appropriate the relation of this work to later developments. I begin with a brief survey of the prehistory of work on singular Lagrangians, followed by some comments on the life of Peter Bergmann. These are included in part to commemorate Peter in this first History of General Relativity meeting following his death in October, 2002.  Then I will address what I perceive to be the principle innovations of his early Syracuse career. Josh Goldberg covered some of this ground in his 2002 report,\cite{goldberg05} but I hope to contribute some new perspectives. I conclude with a partial list of historical  issues that remain to be explored.

\section{Singular Lagrangian prehistory}

All attempts to invent a Hamiltonian version of singular Lagrangian models are based either explicitly or implicitly on Emmy Noether's remarkable second  theorem of 1918.\cite{noether18} I state the theorem using the notation for variables employed by Bergmann in his first treatment of singular systems published in 1949.\cite{bergmann49a} Denote field variables by $y_A$ $(A = 1, \cdots , N)$, where $N$ is the number of algebraically independent components, and let $x$ represent coordinates. Noether assumes that $n$ is the highest order of derivatives of $y_A$  appearing in the Lagrangian, $L(y_A, y_{A, \mu}, \cdots , y_{A, \mu_1 \cdots \mu_n})$, but I will assume that $n = 1$. The extension of the theorem to higher derivatives is straightforward. Then for an arbitrary variation $\delta y_A(x)$ we have after an integration by parts the usual identity
\beq
L^A \delta y_A \equiv \delta L + \frac{\partial}{\partial x^\mu} \left(\frac{\partial L}{\partial y_{A,\mu}} \delta y_A \right), \label{iden}
\eeq
where the Euler-Lagrange equations are
\beq
L^A =: \frac{\partial L}{\partial y_A} -\frac{\partial}{\partial x^\mu} \left(\frac{\partial L}{\partial y_{A, \mu}}\right) = 0.
\eeq
Now suppose that the action is invariant under the infinitesimal coordinate transformation $x'^\mu = x^\mu + \xi^\mu(x)$. Invariance is defined by Noether as follows:
\beq
\int_{\cal R'}L(y'_A, y'_{A, \mu}) d^4\! x' = \int_{\cal R}L(y_A, y_{A, \mu}) d^4\! x. \label{inv}
\eeq
(The notion of invariance was extended later, as we shall see below, to in include a possible surface integral).
Crucial in this definition is the fact the Lagrangian is assumed not to have changed its functional form, guaranteeing that this transformation does not change the form of the equations of motion, i.e., it is a symmetry transformation. Noether writes $ \delta y_A(x) := y'_A(x') - y_A(x) $, and therefore $y'_A(x) =  y_A(x - \xi) + \delta y_A(x)$. She then defines
\beq
\bar \delta y_A (x) := y'_A(x) - y_A(x) = \delta y_A(x) - y_{A , \mu}(x) \xi^\mu(x).
\eeq
This $\bar \delta$ notation was appropriated by Bergmann in his 1949 paper, and retained throughout his life. It is, of course, the Lie Derivative with respect to the vector field $-\xi^\mu$, a terminology introduced by Sledbodzinski in 1931.\cite{sledbodzinski31}  Returning to the elaboration of Noether's theorem, using this notation we may rewrite the invariance assumption (\ref{inv}) as 
\beq
  \bar \delta L \equiv -   \frac{\partial }{\partial x^\mu}\left( L \xi^\mu \right), \label{inv}
 \eeq
so that under a symmetry transformation the identity (\ref{iden}) becomes
\beq
L^A\bar  \delta y_A \equiv \frac{\partial}{\partial x^\mu} \left(\frac{\partial L}{\partial y_{A,\mu}} - L \xi^\mu \right) . \label{div1}
\eeq
Next, let us assume that $\bar \delta$ variations of $y_A$ are of the form 
\beq
\bar \delta y_A ={}^0\! f_{A i}(x, y, \cdots) \xi^i(x) + {}^1\!f_{A i}^\nu(x, y, \cdots) \xi^i_{,\nu} (x), \label{deltay}
\eeq
where we have admitted the possibility of additional non-coordinate gauge symmetry by letting the index $i$ range beyond 3.
We are finally in position to state (and prove) Noether's second theorem: Perform an integration by parts on the left hand side of (\ref{div1}) using (\ref{deltay}), then it follows  for functions $\xi^\mu$ that vanish on the integration boundary that
\beq
L^A \, {}^0\! f_{A \mu} - \frac{\partial}{\partial x^\nu} \left(  L^A \, {}^1\!f_{A \mu}^\nu \right) \equiv 0. \label{bianchi1}
\eeq
In vacuum general relativity these are the contracted Bianchi identities.The derivation from general coordinate symmetry had already been anticipated by Hilbert in 1915 in a unified field-theoretic context.\cite{hilbert15} Weyl applied a similar symmetry argument in 1918.\cite{weyl18} He adapted Noether's theorem to a gravitational Lagrangian ${\cal L_W}$ from which a total divergence has been subtracted so that the highest order of derivatives appearing are $g_{\mu \nu , \alpha}$ . ${\cal L_W}$ is no longer a scalar density, but the extra divergence term can easily be incorporated in its variation,
\beq
{\cal L_W} =\sqrt{-g} R -\left(\sqrt{-g} g^{\mu \nu} \Gamma^\rho_{\mu \rho} -\sqrt{-g} g^{\mu \rho} \Gamma^\nu_{\mu \rho} \right)_{,\nu} = \sqrt{-g} g^{\mu \nu} \left( \Gamma^\sigma_{\rho \sigma} \Gamma^\rho_{\mu \nu} - \Gamma^\sigma_{\mu \rho} \Gamma^\rho_{\nu \sigma} \right).
\eeq
Bergmann and his collaborators will later work with this Lagrangian. It appears in his 1942 textbook.\cite{bergmann42}
Pauli  in 1921 applied similar symmetry arguments, citing Hilbert and Weyl, but curiously never mentioning Noether.\cite{pauli21} Pauli is an important link in this story. Leon Rosenfeld writes in 1930 in his groundbreaking paper on constrained Hamiltonian dynamics that it was Pauli who suggested to him a method for constructing a Hamiltonian procedure in the presence of identities.\cite{rosenfeld30} Rosenfeld did indeed make astounding progress in constructing a gravitational Hamiltonian. Full details will be reported elsewhere,\cite{rossalis} but its relevance specifically to the work of the Syracuse ``school'' will be addressed below.

\section{A brief Bergmann biography}

Peter Bergmann was born in 1915 in Berlin Charlottenburg. His mother, Emmy (Grunwald) Bergmann was one of the first female pediatricians in Germany. She was also the founder in Freiburg in 1925, where she moved with her son and daughter in 1922, of the second Montessori school in Germany. She had taken a course in Amsterdam in the winter of 1923/24 with Maria Montessori. The chemist Max Bergmann, Peter's father, was a student and collaborator of the 1902 Nobel prize winner in chemistry, Emil Fischer. He was appointed the first Director in 1923 of the Kaiser Wilhelm Institut f\"ur Lederforschung in Dresden. He was removed from this position by the new Hitler regime in 1933, despite the personal intervention of the then president of the Kaiser Wilhelm Gesellschaft, Max Planck. He then assumed a position at what was to become Rockefeller  University in New York City in 1936. Max Bergmann is recognized as one of the founders of modern biochemistry. Peter's aunt, Clara Grunwald, was the founder of the Montessori movement in Germany. He had fond memories of visits with his mother's eldest sister in Berlin.\cite{memory} He clearly had benefited from Montessori methods, as attested by his aunt in  references to him in letters written by her from what had become a forced labor camp near F\"urstenwald just outside of Berlin.\cite{grunwald} Clara Grunwald perished with her students in Auschwitz in 1943.

After completing his doctorate in physics at Charles University in Prague in 1936 Peter Bergmann was appointed an assistant to Albert Einstein at the Institute for Advanced Study in Princeton. He worked with Einstein on unified field theory until 1941.  There followed brief appointments at Black Mountain College, Lehigh University, Columbia University and the Woods Hole Oceanographic Institute. In 1947 he joined Syracuse University where he remained until his retirement in 1982. He remained active for many years with a research appointment at New York University. Syracuse became the center for relativity research in the United States in the 1950's and early 1960's, bringing virtually all the leading relativists in the world for either brief or prolonged interaction with Bergmann and his collaborators.  Bergmann concentrated from the beginning on the conceptual and technical challenges of attempts to quantize general relativity. Not unlike Einstein himself, his deep physical intuition was founded on hands-on laboratory experience, in his case extending back to ``enjoyable'' laboratory courses in physics and chemistry as an undergraduate in 1932 at the Technical University in Dresden.\cite{bern} Later on he expressed appreciation for the opportunity that teaching at the graduate level had given him to explore domains outside of relativity. His two-volume set of lectures on theoretical physics are magisterial lucid surveys of the field\cite{bergmann49b}, and it is lamentable that they are now out of print. In fact, the pure mathematical aspect of relativity was not for him especially appealing, and he tended not to work closely with visitors in the 1960's who approached the subject from this perspective.\cite{schucking} For additional biographical material see my short sketch\cite{salisbury05}, and a longer discussion by Halpern. \cite{halpern05}

\section{1949 - 1951}

Bergmann's aim in his 1949 paper is to establish the general classical canonical framework  for dealing with a fairly narrow set of generally covariant dynamical systems, but a set that includes as a special case general relativity described by the Lagrangian ${\cal G}$ above. He assumes that under the infinitesmal  general coordinate transformation $x'^\mu = x^\mu + \xi^\mu(x)$  the $\bar \delta $ transformations are given by
\beq
\bar \delta y_A = F_{A \mu}{}^{B \nu} \xi^\mu_{,\nu} - y_{A, \mu} \xi^\mu, \label{deltaby}
\eeq
where the $F_{A \mu}{}^{B \nu}$ are constants. Noether is not cited in this paper, surely because at this time her theorem was common knowledge.\cite{bergmann62} A principal concern from the start is with the group structure of these symmetry transformations, and with the requirement that canonically realized variations faithfully reproduce the $\bar \delta $ variations.

Due to the intended use of the Lagrangian ${\cal G}$ an additional term will appear on the right hand side of the invariance assumption (\ref{inv}). This eventuality is accommodated by Bergmann with the assumption that $\bar \delta L \equiv  Q^\mu_{, \mu}.$ Rather than consider $\xi^\mu$ that vanish on the integration boundaries, he equivalently requires the identical vanishing of that contribution to the duly rewritten (\ref{div1}) that cannot be written as a total divergence. Thus he obtains the generalized contracted Bianchi identity  (\ref{bianchi1}) that for the variations (\ref{deltaby}) takes the form
\beq
\left( F_{A \mu}{}^{B \nu} y_B L^A \right)_{,\nu} + y_{A, \mu} L^A \equiv 0. \label{bianchi2}
\eeq
It is at this stage that new information is mined from the invariance of the Lagrangian (although without Bergmann's knowledge Leon Rosenfeld had employed similar arguments in 1930)\cite{rosenfeld30}. Since the $\xi^\mu$ are arbitrary functions, the coefficients of all distinct derivatives of $\xi^\mu$ in (\ref{bianchi2}) must vanish identically. In particular the coefficient of  $\frac{\partial^3 \xi^\mu}{\partial (x^0)^3} =: \dddot \xi^\mu $ must be zero,
\beq
F_{A \mu}{}^{B 0} \Lambda^{A C} y_B \equiv 0, \label{null}
\eeq
where
\beq
\Lambda^{A C} :=-  \frac{\partial^2 L}{\partial \dot y_A \partial \dot y_B},
\eeq
is the Legendre matrix.  Thus the Legendre matrix possesses null vectors. This is the signature of singular Lagrangians. 

Bergmann deduces several interrelated consequences. Firstly, since by assumption the Euler-Lagrange equations are linear in $\ddot y_A$, with the linear term of the form 
$\Lambda^{A C} \ddot y_C$, the following four linear combination of equations of motion do not contain accelerations:
\beq
y_B F_{A \mu}{}^{B 0} L^A = 0.
\eeq
Therefore the evolution from an initial time will not be uniquely fixed through an initial choice of $y_A$ and $\dot y_A$.
Secondly, it will not be possible to solve for velocities in terms of canonical momenta $\pi^A := \partial L/ \partial \dot y_A$ since the matrix $\Lambda^{A B}$ cannot be inverted. Thirdly, since 
\beq
y_B F_{A \mu}{}^{B 0} \frac{\partial \pi^A }{\partial \dot y_C} \equiv\frac{\partial  }{\partial \dot y_C} \left( y_B F_{A \mu}{}^{B 0}\pi^A \right) \equiv   0,
\eeq
straightforward integration yields a   constraining relation amongst the momentum $\pi^A$ and configuration variables $y_B$.

Although the central stated objective of this first paper was to prepare the ground for a full-scale quantization of the gravitational field, Bergmann did note that the canonical phase space approach offered a potential new method for solving the classical particle equation of motion problem. Indeed, he expressed a hope shared by Einstein that through avoiding singular field sources, the locations of point particles, it might be possible to eliminate singularities in an eventual quantum gravitational field theory. This hope led in the second paper (BB49), co-authored with Brunings, to the introduction of a parameterized formalism in which spacetime coordinates $x^\mu$ themselves became dynamical variables.\cite{bergbrun49} For the further development of the constrained dynamical formalism this was an unnecessary computational complication. Yet several important results were obtained. In the parameter formalism the Lagrangian is homogeneous of degree one in the velocities. Consequently the Hamiltonian density ${\cal H}$ vanishes identically. It was possible to find immediately seven of the functions of the $y_a$ and conjugate momenta $\pi^b$ whose vanishing follows from the Legendre map $\pi^a(y, \dot y) := \partial L/\partial \dot y_a$. (The range of the index $a$ has been expanded by four to accommodate the spacetime coordinates.)  BB49  recognized that the pullback of the Hamiltonian under the Legendre map yielded a null vector of the Legendre matrix,
\beq
0 \equiv \frac{\partial }{\partial \dot y_a} {\cal H}\left(y, \pi(y, \dot y)\right)  = \frac{\partial {\cal H}}{\partial y_b} \Lambda^{b a}.
\eeq
But the homogeneity of the Lagrangian implies that the velocities are also components of a null vector. It follows that one may set $\dot y_a = \partial {\cal H}/ \partial \pi^a$.  Dirac would soon reach the same conclusion in his parameterized flat spacetime models.\cite{dirac50} Apparently unbeknownst to any of the parties, Rosenfeld had already shown in 1930 that a relation of this form more generally reflects the freedom to alter the velocities without affecting the momenta, albeit in models with Lagrangians quadratic in the velocities.\cite{rosenfeld30} Next, considering variations of ${\cal H}$ independent variations at a fixed time, and using the Euler -Lagrange equations, Bergmann and Bruning obtain the  ``usual''  additional Hamiltonian dynamical equations  $\dot \pi^a =- \frac{\partial {\cal H}}{\partial y_a}$. BB49 do note that there is considerable freedom in the choice of the vanishing Hamiltonian. Given any ${\cal H}$ resulting from the homogeneity of the Lagrangian, one may multiply by an arbitrary function of the spacetime coordinates, or add arbitrary linear combinations of the remaining seven constraints without altering the canonical form of the Hamiltonian equations. They do appear to claim, erroneously however, that the vanishing of all of these possibilities is preserved under the evolution of a fixed Hamiltonian. Unfortunately this renders untenable the proposed Heisenberg picture quantization in which the quantum states are annihilated by all of the constraints ${\cal G}$.

In this paper we find the first statement of the requirement of projectability under the Legendre transformation from configuration-velocity to phase space. Only those functions $\Psi$ that are constant along the null directions $u_a$ of $\Lambda^{a b}$ have a unique counterpart in phase space since
\beq
u_a \frac{\partial }{\partial \dot y_a} \Psi(y, \pi\left(y, \dot y)\right)  = u_a \frac{\partial \Psi}{\partial p^b} \Lambda^{b a} = 0.
\eeq
This requirement remained a concern until it appeared to have been resolved in 1958 through the elimination of lapse and shift variables, as described below. Only much later was the relevance to the canonical symmetry group understood.\cite{leeW90, pss97}

The explicit expression for the Hamiltonian was obtained by Bergmann, Penfield, Shiller, and Zatkis in following paper (BPSZ50).\cite{bergPSZ50} Because of  the complications ensuing in the parameterized formalism, the solution was a daunting task. The work focuses on an algorithm for transforming the Legendre matrix into a so-called ``bordered'' form in which the final eight rows and columns are zero. We will not address the details here since much of the technology was rendered superfluous by the discovery by Penfield, one of Bergmann's students, that the parameterization could be profitably dispensed with. Josh Goldberg vividly recalls the excitement, and it was he who communicated the news to their approving mentor.\cite{goldberg50} Penfield worked with a quadratic Lagrangian of the form
\beq
L = \Lambda^{A \rho B \sigma}(y)  y_{a , \rho} y_{B , \sigma},
\eeq
so
\beq
\pi^A = 2 \Lambda^{A 0 B a} y_{B, a} + \Lambda^{A B} \dot y_B, \label{pi}
\eeq
where
$ \Lambda^{A B} := 2 \Lambda^{A 0 B 0} $ is the Legendre matrix.\cite{penfield51} His task was to find the appropriate linear combination of the $\dot y_A$ such that $\Lambda^{A B}$ is transformed into a bordered matrix. In somewhat more technical terms, he sought a linear transformation in the tangent space of the configuration velocity space such that each null vector acquires a single non-vanishing component. This procedure had already been undertaken by BPSZ50, but its implementation in this context was much simpler. 

Indeed, it is immediately clear from (\ref{pi}) that once a particular solution for $H$ is found, resulting in a fixed $\dot y_A$, any linear combination of the remaining constraints may be added to $H$ since, as also noted by BPSZ, the additional terms do not change $\pi^B$. (Recall that the gradients of constraints with respect to momenta are null vectors.)

As pointed out already by BB49, additional gauge symmetry can easily be incorporated into the formalism, resulting in as many new constraints as there are new gauge functions. Thus both BB49 and BPSZ50 produced Hamiltonians for gravity coupled to electromagetism.

At some time in 1950 the Syracuse group became aware of the pioneering work of Leon Rosenfeld. Reference to Rosenfeld appears in a 1950 Proceedings abstract.\cite{berg50}
James Anderson thinks it is possible that he brought the work to Bergmann's attention\cite{anderson}, and Bergmann showed the paper to Ralph Schiller. In fact, according to Schiller the paper inspired his doctoral thesis.\cite{schiller} In any case, the culminating paper of this period by Bergmann and Anderson (BA51) was written after the discovery, and it does appear that the authors were motivated by it to broaden the final scope of their published investigations of constrained Hamiltonian dynamics. In particular, in addition to abandoning the parametrized theory, BA51 contemplated more general symmetry transformations, similar to those of Rosenfeld, 
\beq
\bar \delta y_A ={}^0\! f_{A i}(x, y, \cdots) \xi^i(x) +\cdots + {}^P\!f_{A i}^{\nu_1 \cdots \nu_P}(x, y, \cdots) \xi^i_{,\nu_1 \cdots \nu_P} (x). \label{deltay}
\eeq
The BA51 collaboration was a watershed in which most of the basic elements of the formalism were completed.  For the first time in this paper the question was asked whether coordinate-transformation-induced variations of the momenta are realizeable as canonical transformations. BA51 assumed that the canonical generator density ${\cal C}$ of these symmetry transformations could be written as
\beq
{\cal C} = {}^0\! A_i \xi^i + {}^1\! A_i \frac{\partial \xi^i}{\partial t} + \cdots + {}^P\! A_i \frac{\partial^P  \xi^i}{(\partial t )^P},
\eeq
where the ${}^M\!A_i$ are phase space functions. Thus it was necessary to show, as they did, that the momenta variations did not depend on time derivatives of $\xi$ of order higher that P; the potential offending term in $\bar \delta \pi^A$ is $2 \Lambda^{A B} (-1)^P\, {}^P\!f_{B i}^{\nu_1 \cdots \nu_P} \frac{\partial  \xi^i}{(\partial t )^P}$, but $\Lambda^{A B}\, {}^P\!f_{B i}^{\nu_1 \cdots \nu_P} $ vanishes identically since it is the coefficient of $ \frac{\partial^{P+1}  \xi^i}{(\partial t )^{P+1}}$ in the generalization of the identity (\ref{bianchi1}). Most importantly, BA51 argued that since that commutator of transformations generated by ${\cal C}'s$ must be of the same form, the ${}^M\!A_i's$ must form a closed Poisson bracket algebra. Furthermore, they were able to show that the ${}^P\!A_i$ are the constraints that follow from the momenta definitions. For these they introduced the term ``primary constraint''. They showed that in order for these constraints to be preserved under time evolution, all of the ${}^M\!A_i's$ were required to vanish; again according to their terminology, ${}^{P-1}\!A_i$ is a secondary constraint, ${}^{P-2}\!A_i$ tertiary, etc. The argument employed here is similar to the one used by Rosenfeld. Up to this point Rosenfeld's results are similar. He does not, however, take the next step in which BA51 derive a partial set of Poisson relations among the ${}^M\!A_i's$. All of these results are displayed explicitly for gravity and a generic generally covariant model that includes Einstein's gravity as a special case.

\section{Preview of some later developments}

It is not possible to do justice to Bergmann's complete oevre in constrained Hamiltonian dynamics in this talk. I will just briefly mention two important developments that will be treated in detail elsewhere, and I will conclude with a teaser of contemporary importance. Much effort was expended in Syracuse in the 1950's in constructing gravitational observables, functions of canonical variables that are invariant under the full group of general coordinate transformations. In 1958 Paul Dirac published his  simplified gravitational Hamiltonian, achieved through a subtraction from the Lagrangian that resulted in the vanishing of four momenta.\cite{dirac58} He argued that the corresponding configuration variables, the lapse and shift functions, could then simply be eliminated as canonical variables. There remained a puzzle over the precise nature of the canonical general coordinate symmetry group. Bergmann and Komar made considerable headway in describing this group in 1972.\cite{bergK72} They showed in particular that the group must be understood as a transformation group on the metric field. This view was forced by the observation that the group involved a compulsory dependence on the metric, and it was manifested in part by the appearance of metric components in the group Poisson bracket algebra. There exists a close relation between these developments and the so-called ``problem of time'' in general relativity. Are invariants under the action of the group necessarily independent of time? The issue is addressed in an early exchange between Bergmann and Dirac, with which I will close.\cite{bergdirac}  In a letter to Dirac dated October 9, 1959 Bergmann wrote ``When I discussed your paper at a Stevens conference yesterday, two more questions arose, which I should like to submit to you: To me it appeared that because you use the Hamiltonian constraint $H_L$ to eliminate one of the non-substantive field variables ${\it K}$, in the final formulation of the theory your Hamiltonian vanishes strongly, and hence all the final variables, i. e. $\tilde e^{r s}, \tilde p^{r s}$, are ``frozen'', (constants of the motion). I should not consider that as a source of embarrasment, but Jim Anderson says that in talking to you he found that you now look at the situation a bit differently. Could you enlighten me? '' Here is Dirac's response, dated November 11, 1959: ``If the conditions that you introduce to fix the surface are such that only one surface satisfies the condition, then the surface cannot move at all, the Hamiltonian will vanish strongly and the dynamical variables will be frozen. However, one may introduce conditions which allow an infinity of roughly parallel surfaces. The surface can then move with one degree of freedom and there must be one non-vanishing Hamiltonian that generates this motion. 
I believe my condition $g_{rs} p^{rs}$ is of this second type, or maybe it allows also a more general motion of the surface corresponding roughly to Lorentz transformations. The non-vanishing Hamiltonian one would get by subtracting a divergence from the density of the Hamiltonian.

\section*{Acknowledgements}
I would like to thank the Instituto Orotava for its hospitality and the Max Planck Institute f\"ur Wissenschaftsgeschichte for inviting me to contribute to this meeting. Thanks also to Josh Goldberg for his critical reading of this paper and helpful comments.

 \end{document}